# Predictive and Spatially Aware Scheduling in Flexible Duplexing for Deterministic Communications

Syed Morsleen Riaz, Baldomero Coll-Perales, M. Carmen Lucas-Estañ, Javier Gozalvez, Miguel Sepulcre
Networked Systems Laboratory, Universidad Miguel Hernández de Elche, Elche (Alicante), Spain
sriaz@umh.es, bcoll@umh.es, m.lucas@umh.es, j.gozalvez@umh.es, msepulcre@umh.es

***Abstract*—Next generation wireless networks must sustain deterministic service levels for time-sensitive closed-loop applications. Flexible duplexing (FD) is an efficient solution to support these services, as it enables simultaneous uplink (UL) and downlink (DL) transmissions over orthogonal resources within the same band. However, simultaneous UL and DL transmissions can create conflicts that degrade performance due to interference from in-band emissions (IBE) and UL-to-DL cross-link interference (CLI). In this paper, we propose to use traffic forecasting and predictive scheduling to mitigate UL/DL conflicts in FD. Our proposal exploits traffic predictions to increase the likelihood of scheduling CLI-free UL and DL transmissions, and leverages spatial diversity to minimize the impact of unavoidable conflicts. Results show that the proposed scheme reduces UL/DL scheduling conflicts and improves the SINR of conflicted transmissions by more than 5 dB. This leads to gains of over 40% in the number of successfully completed transmissions compared to reference FD schemes.**

## I. Introduction

Beyond-5G/6G wireless networks must support increasingly heterogeneous traffic demands, including deterministic and time-sensitive services required by safety-critical control applications such as digital twins and sensor-actuator interactions. To support the dynamic and varying uplink (UL) and downlink (DL) traffic demands of such services, full duplexing has been investigated as a potential solution. In-band full duplexing enables simultaneous UL and DL transmissions over the same time-frequency resources, but it is challenged by the complex processing needed to cancel its self-interference [1]. Sub-Band Full-Duplexing (SBFD) simplifies the full duplexing operation and removes self-interference by partitioning the band into sub-bands dedicated to either UL or DL. Flexible duplexing (FD) provides additional versatility by allowing UL and DL transmissions to dynamically use orthogonal time-frequency resources, without being restricted to fixed subbands [2].

FD offers a favorable complexity-flexibility tradeoff for next-generation wireless networks (3GPP TR38.858 v18.2.0). This paper studies, for the first time to the best of our knowledge, the capacity of FD to support time-sensitive closed-loop services. To this end, we propose a novel scheduling scheme aimed at exploiting the scheduling flexibility granted by FD while dealing with the inherent cross-link interference (CLI) caused by the simultaneous UL and DL transmissions. We focus on UL-to-DL interference, as DL-to-UL interference is negligible in single-cell scenarios [2]. Prior work has proposed managing such interference by employing power control mechanisms or guard bands to reduce energy leakage between UL and DL transmissions [3]. Guard bands reduce spectral efficiency, while power control requires accurate channel estimation which might incur non-negligible overhead [3]. Predictive knowledge about future traffic demands can also improve scheduling decisions in scenarios with mixed traffic and diverse communication requirements [4][5]. In this context, our proposal jointly considers the traffic forecasts of upcoming demands and spatial diversity of the devices generating such traffic for minimizing CLI. By anticipating future traffic demands, the scheduler can prioritize the simultaneous scheduling of UL and DL transmissions from devices that are highly separated from each other. This reduces their CLI level and increases the likelihood that both transmissions are successfully completed.

We illustrate the motivation of our proposal through a 6G in-vehicle use case part of 3GPP TR 22.870 (v1.0.0). Following the automotive shift towards zonal E/E architectures with centralized computing nodes [6], the scenario considers an in-vehicle local network deployed to support closed-loop interactions between on-board sensors and actuators, managed by a central decision-making entity. The traffic generated by the on-board sensors is derived from realistic driving environments using a Level 3 (L3) autonomous driving platform. Evaluation results against reference FD schemes demonstrate that considering traffic forecasting and spatial diversity increase the likelihood of achieving CLI-free scheduling of UL and DL transmissions from different closed-loop sessions. When simultaneous UL and DL transmissions are unavoidable, our proposal improves the SINR of received packets by more than 5dB. Altogether, this leads to an increase of up to 43% in the number of successful closed-loop sessions compared to scheduling schemes that do not leverage spatial diversity or predictive knowledge of future traffic demands.

## II. System Model

We consider a 6G local wireless network or subnetwork comprising *N* sensor nodes with limited capabilities (Sub-Network Elements, SNEs) which are supported by a Higher Capabilities (HC) node acting as a gNodeB [7]. SNEs operate in half-duplex mode and the HC is FD capable. The HC implements a dynamic scheduling (DS) decision-making to support time-sensitive closed-loop traffic with deterministic requirements. SNEs generate UL traffic which is processed and sent back in the DL through the HC. UL and DL use 5G NR-like orthogonal frequency division multiple access (OFDMA) over a shared time-frequency band, and follows the 5G NR frame structure as defined in 3GPP TS 38.300 (v17.11.0). Resource allocation is done at the resource block (RB) granularity. UL and DL are prioritized in dedicated subbands (see Fig. 1), but the HC may also schedule them over the full band to fully exploit FD.

Let $pkt_n^{UL}(t_n)$ denote a packet of size $s_n^{UL}$ bytes generated by $SNE_n$, with $n \in (0, N\text{-}1)$, at time $t_n$ for UL. The $SNE_n$ sends

This work has been partially funded by the European Commission Horizon Europe SNS JU 6G-SHINE (GA 101095738) project, by MCIN/AEI/10.13039/501100011033 (PID2023-150308OB-I00), by Generalitat Valenciana (CIAICO/2024/167), and UMH.

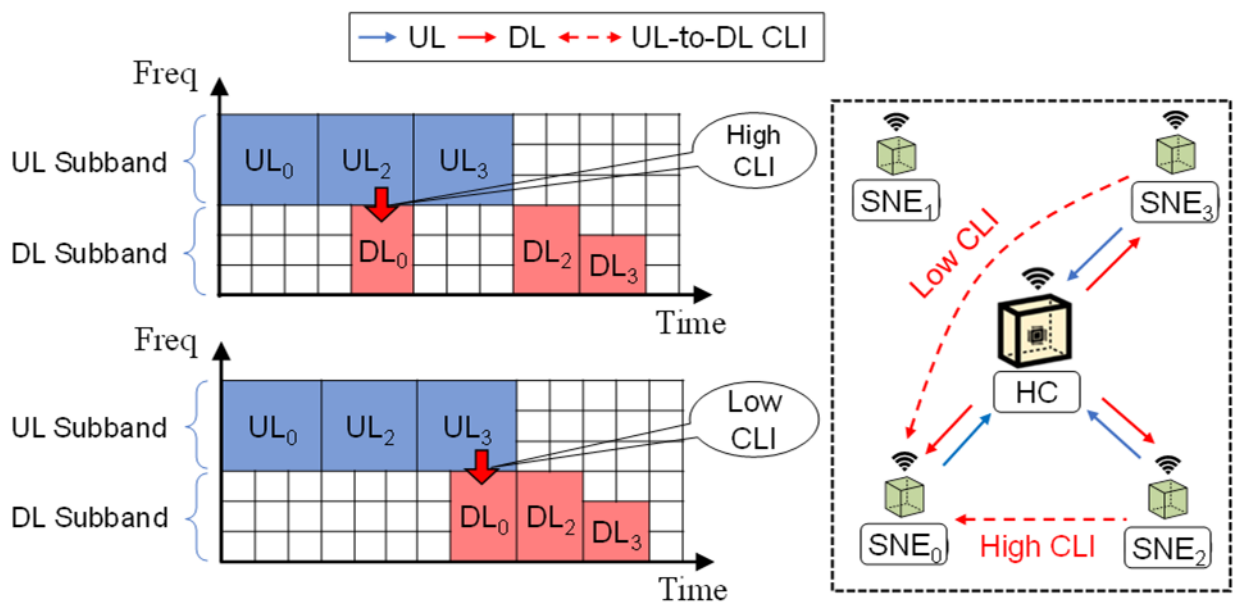

Fig. 1. Illustration of cross-link interference

a scheduling request (SR) to the HC to request RBs for transmitting this packet with the latency requirement $l_n$ of the closed-loop transmission. After UL packet is processed, the resulting $pkt_n^{DL}(t_n)$ is generated for DL. Processing is assumed to take $Y$ slots and DL packet size is $s_n^{DL} = \alpha \cdot s_n^{UL}$; where $\alpha$ is the processing ratio. The HC schedules both UL and DL packets within the deadline $d_n = t_n + l_n$. The RBs required for UL/DL is determined as $R_n^{UL/DL} = f(s_n^{UL/DL}, MCS)$, considering packet size and modulation and coding scheme (MCS) following 3GPP TS 38.300.

Fig. 1 illustrates a scenario using 4 SNEs ($SNE_0 - SNE_3$) communicating with the HC. We highlight two examples of simultaneous UL and DL transmissions (UL/DL conflicts) where in-band emissions (IBE) cause CLI, i.e., UL power leakage interferes with a DL transmission on adjacent resources. This setup illustrates the higher CLI that $SNE_0$ may experience due to its closer proximity to $SNE_2$, compared to the case where $SNE_0$'s DL transmission is scheduled alongside $SNE_3$'s UL, which is a more distant node. These examples highlight the role of spatial diversity in CLI.

IBE is modeled following 3GPP specifications in 3GPP TS 38.300 (v17.11.0) which defines leakage limits. The limit (in dB) of the ratio of the leakage power to DL RBs with respect to the UL power is defined as [2]:

$$R_{leak_{UL \to DL}}[\text{dB}] = max\left\{-25 - 10\,log_{10}\left(\frac{N_{RB}}{L_{CRB}}\right),\ 20\,log_{10}(EVM) - 3 - 5(|\Delta_{RB}| - 1)/L_{CRB},\ -57\,dB + 10\,log_{10}\left(\frac{SCS}{15\,kHz}\right) - P_{UL}\right\}, \quad (1)$$

where $N_{RB}$ is the number of RBs allocated to UL, $L_{CRB}$ is the number of RBs spanning the whole grid, $\Delta_{RB}$ is the gap (in RBs) between UL and DL RBs, $SCS$ is the subcarrier spacing (in kHz), and $P_{UL}$ is the average UL transmission power (in dBm). EVM is the error vector magnitude, which measures the modulation accuracy, i.e., how far the transmitted constellation points deviate from their ideal reference constellations. 3GPP TS 38.101-1 (v18.0.0) defines maximum EVM allowed (in percent) for each modulation scheme, which reflect stricter EVM requirements for higher MCS (e.g., 17.5% EVM for QPSK Vs. 8% EVM for 64QAM). The limit of the leakage power from a SNE transmitting in the UL can then be computed as:

$$\sigma^2_{leak_{UL \to DL}} = \frac{P_{UL}}{r_{leak_{UL \to DL}}}, \quad (2)$$

where $r_{leak_{UL-DL}}$ is the linear value of $R_{leak_{UL-DL}}$ defined in (1). Following [2], the magnitude of the leakage, $\delta_{leakage}$, can be modeled as a Gaussian random variable with zero mean and standard deviation $\sigma_{leak_{UL \to DL}}$, and the instantaneous power of the leakage can be represented as $\delta^2_{leakage}$. Then, the IBE from UL SNE to DL SNE can be computed as:

$$IBE_{UL \to DL} = PL_{UL \to DL} \times \delta^2_{leakage}, \quad (3)$$

**Algorithm I**: PredSpat-FD Scheduling
Input: $pkt_n^{UL}$, $pkt_n^{DL}$, $t_n$, $d_n$, set $\Phi$ of predicted pkts ($\hat{t}_i$, $\widehat{pkt}_i^{UL}$, $\widehat{pkt}_i^{DL}$ $\hat{d}_i$ $\forall I \in \Phi$
Output: RB allocation for $pkt_n^{UL}$ and $pkt_n^{DL}$

1. Define transmission window for each $pkt_i$ in $\Phi$: $w_i = [\hat{t}_i, \hat{d}_i]$
2. Mark RBs already allocated within $w_i$ as as *RB unavailable*
3. Identify overlapping predicted pkts: $\Phi \leftarrow \{t_n < \hat{d}_i \;\&\&\; \hat{t}_i \leq d_n\}$
4. Mark *predicted regions* for all i ∈ Φ
5. Call (*Allocation in free regions*)
6. If no UL/DL conflict-free allocations exists
7. Rank *predicted regions* i ∈ Φ w.r.t *spatial diversity*
8. Call (*Allocation in predicted regions*)
9. if feasible then
10. *allocate* RBs for $pkt_n^{UL}$ and $pkt_n^{DL}$
11. return *allocation*
12. Else
13. fallback *allocate* $pkt_n^{UL}$, then $pkt_n^{DL}$ from start in $w_n$
14. return *allocation*

where $PL_{UL \to DL}$ is the path loss between the SNE transmitting in UL and SNE receiving in DL. The signal-to-interference-plus-noise ratio (SINR) for SNE receiving in the DL can be calculated as:

$$SINR_{DL} = \frac{P_{DL}^{rx}}{IBE_{UL \to DL} + P_{noise}}, \quad (4)$$

where $P_{DL}^{rx}$ is the received power at the SNE scheduled in the DL. $P_{DL}^{rx}$ is computed as $P_{HC}^{tx} \times PL_{HC \to DL}$, where $P_{HC}^{tx}$ is the transmit power of the HC and $PL_{HC \to DL}$ is the pathloss experienced between the HC and the SNE receiving in the DL. $P_{noise}$ is the noise power, modeled as Additive White Gaussian Noise (AWGN). Using the $SINR_{DL}$ in (4), the maximum number of bits that can be reliably transmitted in the DL can be computed as:

$$TR_{DL} = BW_{DL\,SNE} . log_2(1 + SINR_{DL}) . T_s, \quad (5)$$

where $BW_{DL\,SNE}$ is the total bandwidth of the RBs allocated to the SNE receiving DL, and $T_s$ is the slot duration. Then, the DL transmission will be considered successful (i.e., not affected by the UL IBE) if $TR_{DL}$ is greater than or equal to the size of the DL packet.

## III. Proposal Description

We propose a joint prediction-based and spatial-diversity-aware dynamic scheduling scheme for FD (named PredSpat-FD). The scheme aims to minimize UL/DL conflicts and CLI, which challenge the deployment of FD. Without loss of generality, the scheme is presented for time-sensitive closed-loop sessions requiring UL and DL transmissions to meet strict deadline and SINR constraints. UL-to-DL CLI is the main factor degrading SINR in FD. Our proposal leverages traffic predictions to enable conflict-free UL/DL transmissions. When conflicts cannot be avoided, it prioritizes resource allocation based on the spatial separation between the UL transmitter and the DL receiver, in order to reduce their CLI. The operation of the PredSpat-FD scheme is illustrated in Algorithm I and detailed in Fig. 2 and Fig. 3, distinguishing whether current UL and/or DL transmissions are allocated in a region unlikely occupied by upcoming transmissions (Section III.A) or not (III.B). The deployment assumed for this example is the one used in Fig. 1.

### A. Allocation in free regions

Upon receiving the SR from $SNE_0$, PredSpat-FD first identifies the transmission window of its closed-loop session and checks the available RBs within that window (Fig. 2.a). The $SNE_0$'s transmission window is defined by the time $t_0$ at which the UL packet ($pkt_0^{UL}$) is generated and the deadline $d_0 = t_0 + l_0$ at which both $pkt_0^{UL}$ and the DL packet ($pkt_0^{DL}$) must

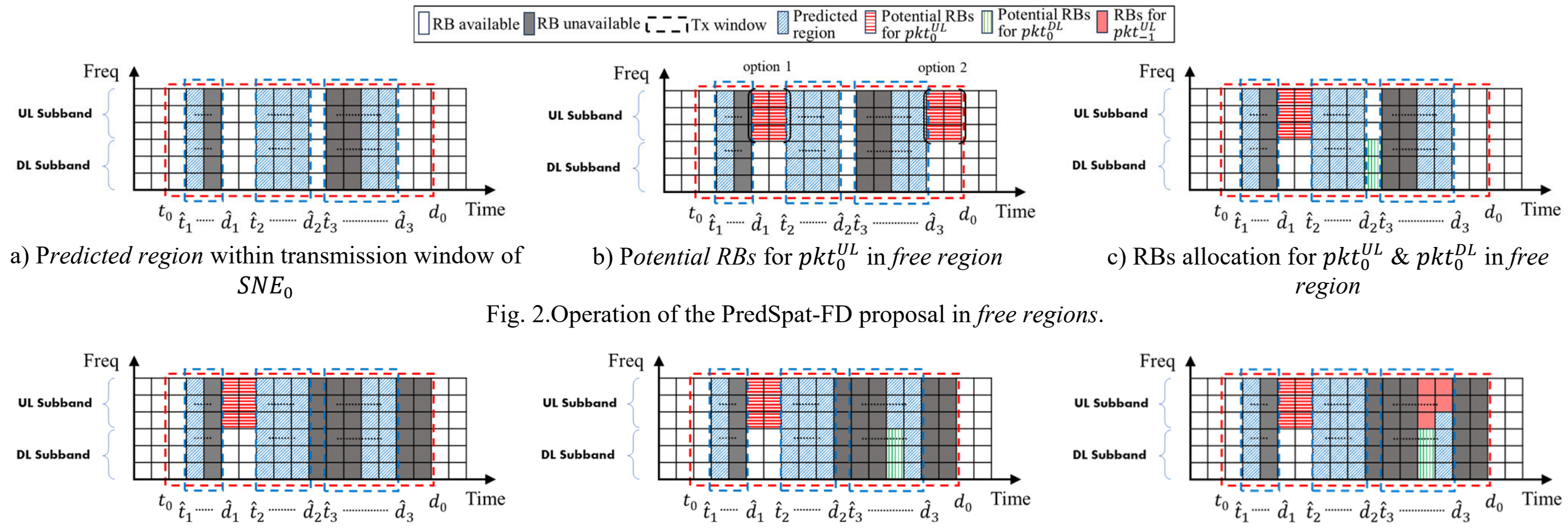

a) *Predicted region* within transmission window of $SNE_0$ b) *Potential RBs* for $pkt_0^{UL}$ in *free region* c) RBs allocation for $pkt_0^{UL}$ & $pkt_0^{DL}$ in *free region*

Fig. 2.Operation of the PredSpat-FD proposal in *free regions*.

a) No available RBs for $pkt_0^{DL}$ in *free region* b) RBs allocation for $pkt_0^{DL}$ in *predicted region* of $SNE_3$ c) Allocation of RBs for $pkt_0^{DL}$ in *predicted region* of $SNE_3$ considering path loss criterion be scheduled (Algorithm I: step 1). Resources previously allocated within the transmission window are marked as *RB unavailable* (Algorithm I: step 2).

Fig. 3. Operation of the PredSpat-FD proposal in *predicted regions*.

The scheduler then determines the set $\Phi$ of predicted packets whose transmission window may overlap with that of the $SNE_0$. The overlap happens when $t_0 < \hat{d}_i$ and $\hat{t}_i \leq d_0$, $\forall\ i\ \varepsilon\ \Phi$ (Algorithm I: step 3). To this aim, a predictor in the scheduler forecasts the generation time $\hat{t}_i$ and size $\hat{s}_i$ of the next *P* packets after $t_0$. As shown in Fig. 2.a, the transmission windows of the predicted packets or *predicted regions* are defined by their predicted generation time $\hat{t}_i$ and deadline $\hat{d}_i = \hat{t}_i + l_i$, $i \in \Phi$ (Algorithm I: step 4). Within each predicted region, $\hat{R}_n^{UL} = f(\hat{s}_n^{UL}, MCS)$ of $\widehat{pkt}_i^{UL}$ and $\hat{R}_n^{DL} = f(\hat{s}_n^{DL}, MCS)$ of $\widehat{pkt}_i^{DL}$ are accounted.

Once the *predicted regions* are marked, PredSpat-FD searches for available RBs to schedule the current transmissions of $SNE_0$ within its transmission window. The scheduler first identifies RBs available in *free regions*, i.e., the portion of the transmission window that does not overlap with any *predicted region* (Fig. 2.b). The objective of this initial search is twofold: (i) to avoid assigning $SNE_0$'s closed-loop session to RBs that may be needed by future SNE's sessions, thereby ensuring that latency requirements are met for both current and upcoming closed-loop sessions; and (ii) to prevent allocations in regions where UL and DL transmissions might be simultaneously allocated (i.e., UL/DL conflicts), which would cause CLI (Algorithm I: step 5). The example in Fig. 2.b presents two options for the initial allocation of $pkt_0^{UL}$ in the UL subband. However, option 2 must be discarded, as it does not leave enough room for the allocation of $pkt_0^{DL}$. Fig. 2.c shows a possible solution for $SNE_0$'s closed-loop session, where the $R_0^{UL}$RBs fit in the UL subband of a *free region* and the $R_0^{DL}$RBs fit in the DL subband of another free region, while maintaining the *Y*-slot gap required between UL and DL transmissions. In a more general case, if the scheduler identifies that multiple valid options exist, it selects the one that minimizes the closed-loop transmission latency.

Other scenarios not illustrated in Fig. 2 may arise in which the available RBs in the UL or DL subband of a *free region* are fewer than $R_0^{UL}$ or $R_0^{DL}$, respectively. In this case, the scheduler can allocate $R_0^{UL}$ and/or $R_0^{DL}$ RBs across the full band, without being restricted to subbands allocations. This option is only considered by the scheduler if UL/DL conflict-free allocations within the subbands are not possible as it limits the benefits of FD; a single transmission monopolizes the full band and prevents possible immediate access for upcoming transmissions.

## B. *Allocation in predicted regions*

Allocation in *predicted regions* is only considered when UL/DL conflict-free allocations in the *free regions* are not possible (Algorithm I: step 6). Fig. 3.a continues the example from Fig. 2.c, assuming that only the current UL transmission of $SNE_0$ could be allocated in a *free region*. Fig. 3.a illustrates the allocation of $pkt_0^{DL}$ is now only possible if it happens within the *predicted regions* of $SNE_2$ (i.e., between $\hat{t}_2$ and $\hat{d}_2$) or $SNE_3$ (i.e., between $\hat{t}_3$ and $\hat{d}_3$). To account for possible CLI, the scheduler evaluates the spatial diversity between $SNE_0$ and $SNE_2$ and $SNE_3$, and ranks them based on their spatial separation (Algorithm I: step 7). Based on the deployment illustrated in Fig. 1, the scheduler considers first the *predicted region* of $SNE_3$ for the allocation of $pkt_0^{DL}$. The allocation is performed in the DL subband, or over the full band if insufficient space is available (as explained in Section III.A), provided that enough resources remain for the upcoming transmission of $SNE_3$. Otherwise, the scheduler would consider the *predicted region* of $SNE_2$. Fig. 3.b shows a possible scenario where the $R_0^{DL}$RBs are allocated in the DL subband within the predicted region of $SNE_3$ (Algorithm I: step 8-11).

The scheduler may need to account for the CLI with previously allocated transmissions within the *predicted regions*. Fig. 3.c illustrates a hypothetical scenario in which the only option for the allocation of $pkt_0^{DL}$ in the predicted region of $SNE_3$ creates an UL/DL conflict with the RBs allocated to $pkt_{-1}^{UL}$ ; subindex -1 denotes any previous transmission from a hypothetical $SNE_{-1}$. In this case, the scheduler would allocate the $R_0^{DL}$RBs associated with $pkt_0^{DL}$ in conflict with the $pkt_{-1}^{UL}$ as long as the spatial separation between $SNE_0$ and $SNE_{-1}$ is larger than the spatial separation between $SNE_0$ and $SNE_3$. Otherwise, it searches for an alternative *predicted region* following the established ranking.

Apart from the scenarios illustrated in Fig. 3, both $pkt_0^{UL}$ and $pkt_0^{DL}$ may need to be allocated in the *predicted regions*. In such a case, the scheduler considers the allocation order by assigning $pkt_0^{UL}$first and then $pkt_0^{DL}$, while maintaining the required *Y*-slot gap between them. If no *predicted regions* in $\Phi$ satisfy the above constraints, the scheduler avoids dropping the current closed-loop session and applies a latency-minimization policy. It allocates $pkt_0^{UL}$ and $pkt_0^{DL}$ in available RBs from the beginning of its transmission window, using subbands first and full-band allocation only if necessary (Algorithm I: step 12-14).

The computational complexity of PredSpat-FD scheme is $O(N \log N + T \cdot F)$, where $T$ denotes the number of time slots within the transmission window and $F$ represents the number of RBs in the frequency domain, and depends on sorting predicted packets and the time–frequency resource grid used for resource allocation.

# IV. Evaluation

## A. Scenario

We consider an in-vehicle 6G subnetwork (3GPP TR 22.870, [7]) comprising 10 SNEs and a central HC unit responsible for radio resource management decision-making. The network is configured with a 5 MHz bandwidth, and uses the standard 5G NR frame structure, with 10 ms radio frames divided into 10 subframes of 1 ms. Numerology (μ) 1 is considered, with a subcarrier spacing (SCS) of 30 kHz, which results in 0.5 ms slots containing 14 OFDM symbols with a normal cyclic prefix. We select a 30 kHz SCS as it is well suited for limited bandwidth operations and aligns with the constraints of low-complexity SNEs. The available 5 MHz band can be flexibly utilized for UL and DL orthogonal transmissions organized in RBs, as described in Section II. The SNEs represent the ADAS sensors (5x cameras and 5x radars) mounted on the top, front, rear and sides of an L3 autonomous vehicle. Path loss (PL) is modeled using in-vehicle field measurements reported in [8], accounting for Line-of-Sight (LoS) and Obstructed-LoS (OLoS) conditions, penetration loss, and noise variations. The resulting PL values, ranging from 70 to 95 dB, are used to characterize the attenuation of the different links based on their separation distances. Both the sensors and the HC transmit at a fixed power of 10 dBm. We evaluate scenarios with different MCS configurations: indices 6 (mod:16QAM, code rate:0.42), 11 (64QAM, 0.46), 15 (64QAM, 0.65), 20 (256QAM, 0.67), and 27 (256QAM, 0.93) as defined in 3GPP TS 38.214 (v16.2.0). The selected MCS determines the number of RBs required for UL/DL transmissions and hence the network load (i.e., resource usage). This setup ensures successful UL transmissions from the SNEs to the HC. However, potential UL-to-DL CLI (i.e., UL/DL conflicts), combined with the selected MCS, may compromise DL transmissions. The DL transmission outcome is determined by the experienced $SINR_{DL}$ and $TR_{DL}$, and the DL packet size (see Section II).

## B. Scheduling schemes and in-vehicle traffic

The proposed PredSpat-FD scheme uses an LSTM-based traffic predictor [5], trained on realistic sensor data from a Connected and Automated Mobility (CAM) platform integrating CARLA and AUTOWARE in an L3 autonomous vehicle. The UL traffic from the SNEs to the HC is derived from CAM traces generated by the cameras and radars during urban driving scenarios. The predictor forecasts the size of the next ten packets, achieving an accuracy of 85% for the first packet, 78% for the fifth, and 70% for the tenth packet [5]. While prediction accuracy can influence the scheduling performance, improving the predictor is beyond the scope of this paper. Flexible duplexing is inherently robust to prediction errors, enabling dynamic use of UL/DL sub-bands or, when needed, the full band for UL or DL transmissions.

To account for different network loads, DL packets from the HC to SNEs are modeled as {20%, 30%, or 50%} of the UL packet size. A minimum separation of three time slots (i.e., 1.5 ms with 30 kHz SCS) is assumed between the reception of a UL packet and the transmission of the corresponding DL packet. The scheduler's effectiveness is evaluated in terms of its ability to successfully complete the closed-loop session within a deadline $d$. Two traffic configurations are defined: Configuration 1 (C1) sets $d$=50 ms deadline for cameras traffic and 10 ms for radars, whereas Configuration 2 (C2) sets 50 ms for cameras and 25 ms for radars.

The proposed PredSpat-FD scheme is benchmarked against two reference scheduling schemes to highlight the benefits of prediction and spatial diversity. The first, Flexible Duplexing with Minimum Latency (MinLat-FD), schedules UL and DL packets with the objective of minimizing closed-loop latency [9]. The second, Predictive Flexible Duplexing (Pred-FD), leverages traffic predictions to allocate UL and DL resources in *free regions* but it does not apply spatial diversity criteria when UL/DL conflicts are unavoidable. Instead, Pred-FD selects the first *predicted region* that has sufficient resources for both the current and predicted transmissions.

## C. Results

Fig. 4 illustrates the percentage of satisfied closed-loop sessions achieved by the three scheduling schemes under evaluation. Results are reported as a function of the MCS used for UL and DL transmissions, and for the two traffic configurations (C1 in Fig. 4.a and C2 in Fig. 4.b). The obtained results show that the proposed PredSpat-FD scheme outperforms the two reference schemes in all evaluated scenarios. The relative gains of the proposed scheme range from 2.34% to 8.72% and 2.74% to 6% with respect to Pred-FD under C1 and C2, respectively, and from 2.02% to 43.85% and 3.68% to 11.40% with respect to MinLat-FD.

Fig. 4 shows higher performance for all schemes with the increasing MCS index. Using higher MCS indices reduces the number of RBs needed for each transmission, which results in less utilization of RBs in the grid and hence more opportunities to schedule conflict-free UL and DL transmissions. This is represented in Fig. 5 that shows the percentage of conflicted UL transmissions relative to the total number of UL transmissions. An UL transmission is classified as conflicted when it is allocated simultaneously with the DL transmission of another SNE's closed-loop session. The obtained results show the decreasing number of UL/DL conflicts as higher MCS indices are used for all evaluated scheduling schemes. MinLat-FD experiences the highest ratio of UL/DL conflicts since it prioritizes minimizing the latency. Although MinLat-FD achieves lower closed-loop session latency (e.g., ~15 ms at 60th percentile vs. ~26 ms for PredSpat-FD with MCS 6), increased UL/DL conflicts reduce the number of satisfied closed-loop sessions (Fig. 4).

Pred-FD and PredSpat-FD focus on completing the closed-loop sessions within their deadlines (without seeking to minimize latency) while leveraging traffic predictions to allocate UL/DL transmissions. The advantages of considering traffic predictions are demonstrated with the lower conflicts experienced by the Pred-FD and PredSpat-FD schemes compared to MinLat-FD (Fig. 5). Both schemes achieve similar reductions in UL/DL conflicts with respect to MinLat-FD across all MCSs under C1 (Fig. 5.a). Under C1, the shorter radar deadline (i.e., 10ms) limits the duration of its transmission windows and increases the availability of free regions, where Pred-FD and PredSpat-FD primarily seek to allocate UL/DL conflict-free transmissions (see Section III.A). This explains the lower number of conflicts under C1 compared to C2 shown in Fig. 5. Under C2, the longer radar deadline (i.e., 25 ms) increases the overlap between radar and camera transmission windows, thereby reducing the availability of free regions. In this setting, the proposed PredSpat-FD scheme achieves further reductions in UL/DL

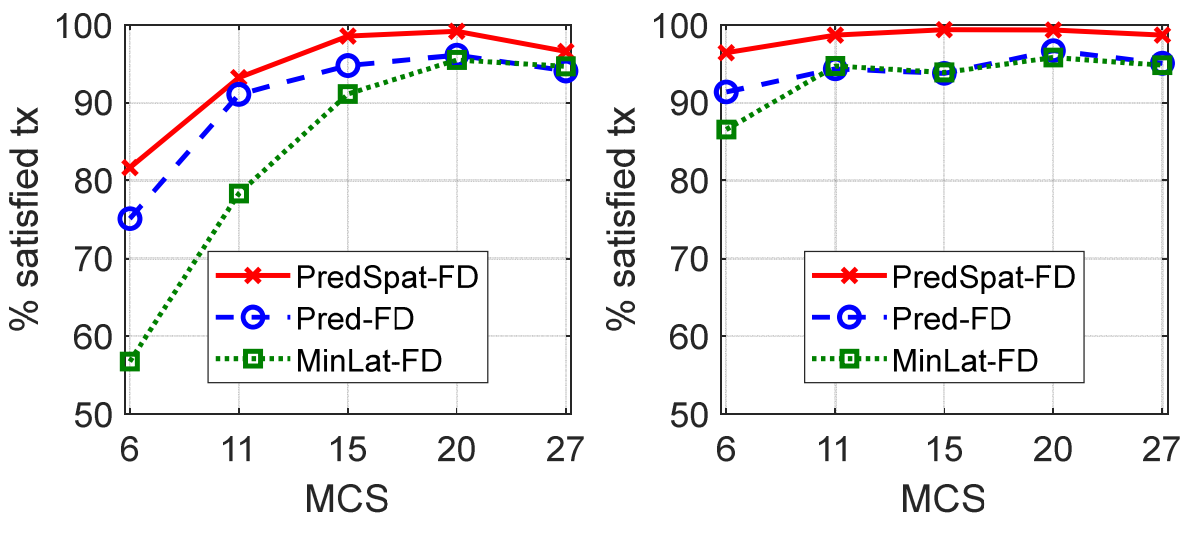

a) Traffic configuration C1. b) Traffic configuration C2.

Fig. 4. Percentage of satisfied closed-loop sessions (DL = 30% of UL).

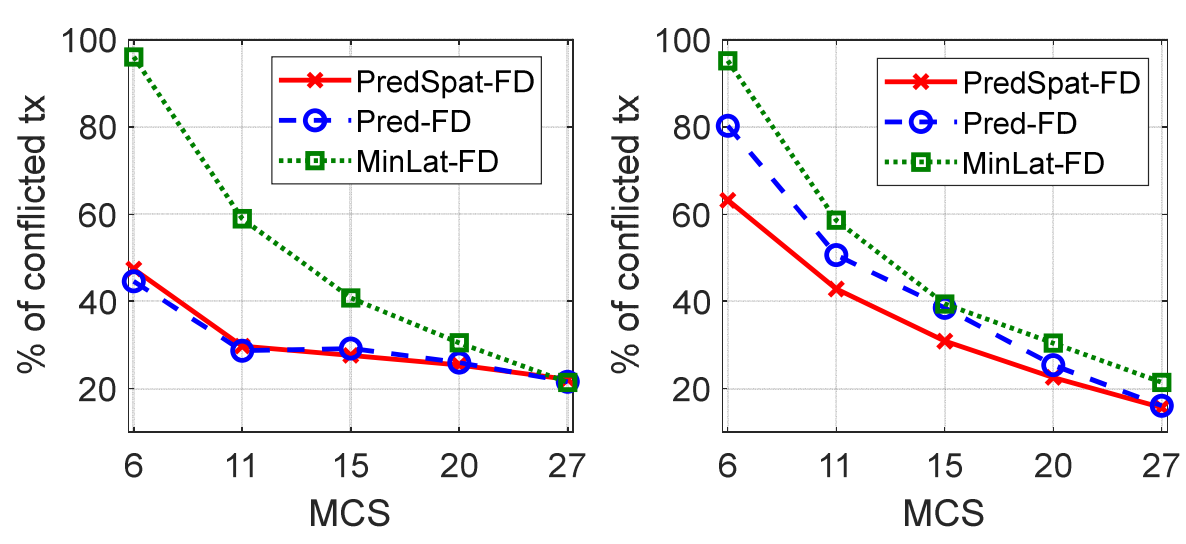

a) Traffic configuration C1. b) Traffic configuration C2.

Fig. 5. Percentage of total conflicted transmissions (DL = 30% of UL).

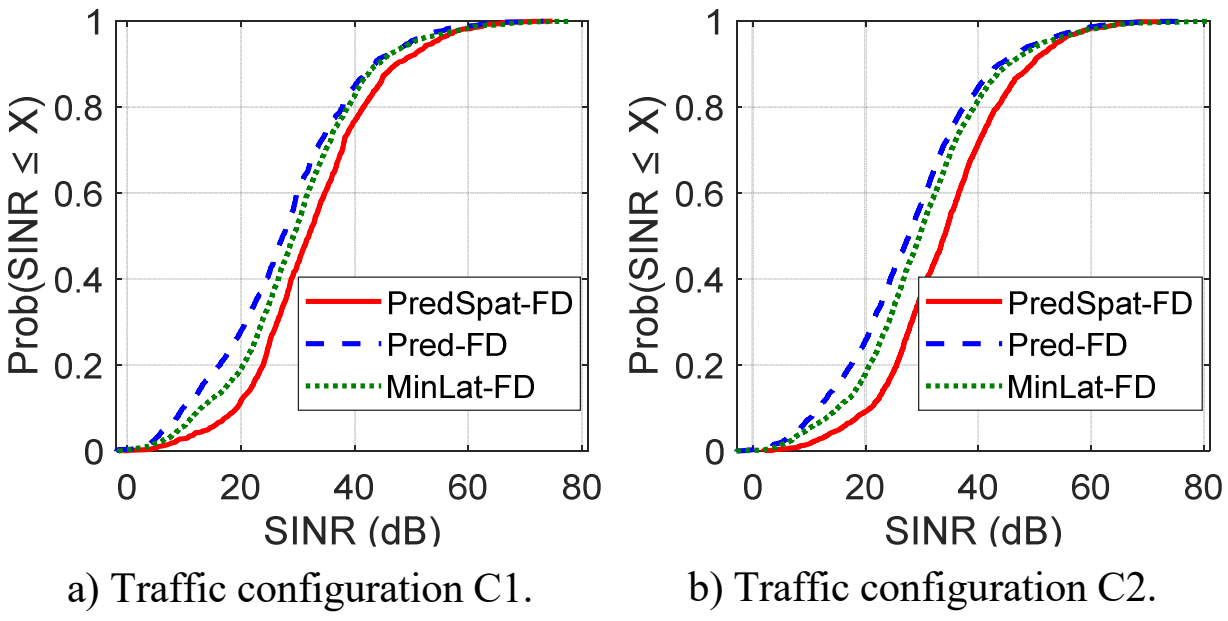

a) Traffic configuration C1. b) Traffic configuration C2.

Fig. 6. CDF of the SINR values for MCS 11 (DL = 30% of UL).

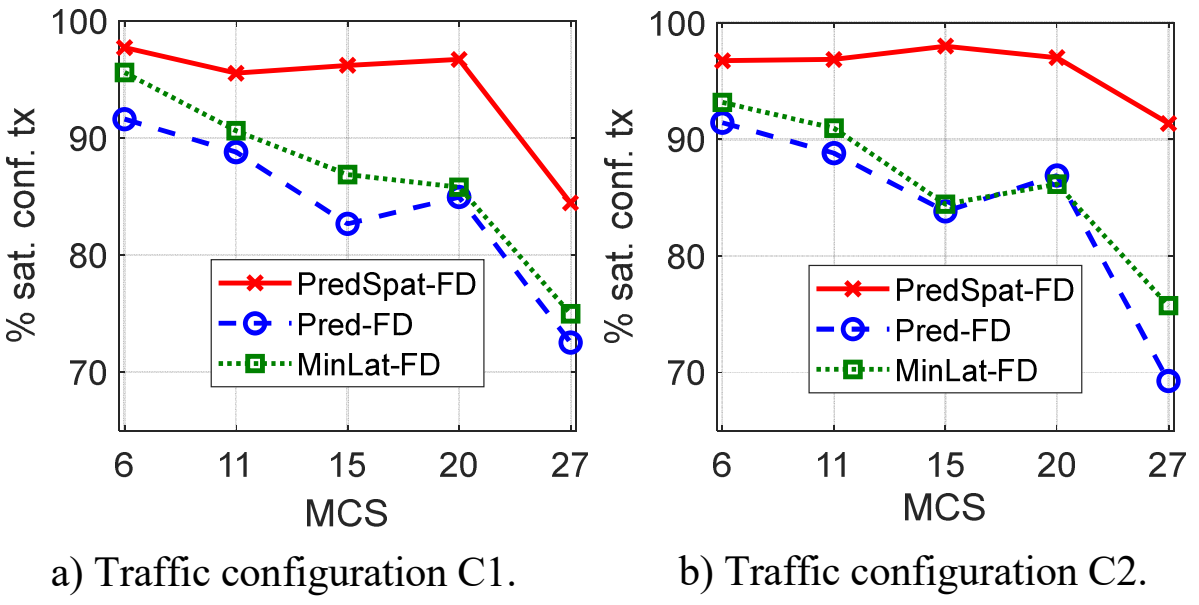

a) Traffic configuration C1. b) Traffic configuration C2.

Fig. 7. Percentage of satisfied conflicted transmissions (DL = 30% of UL).

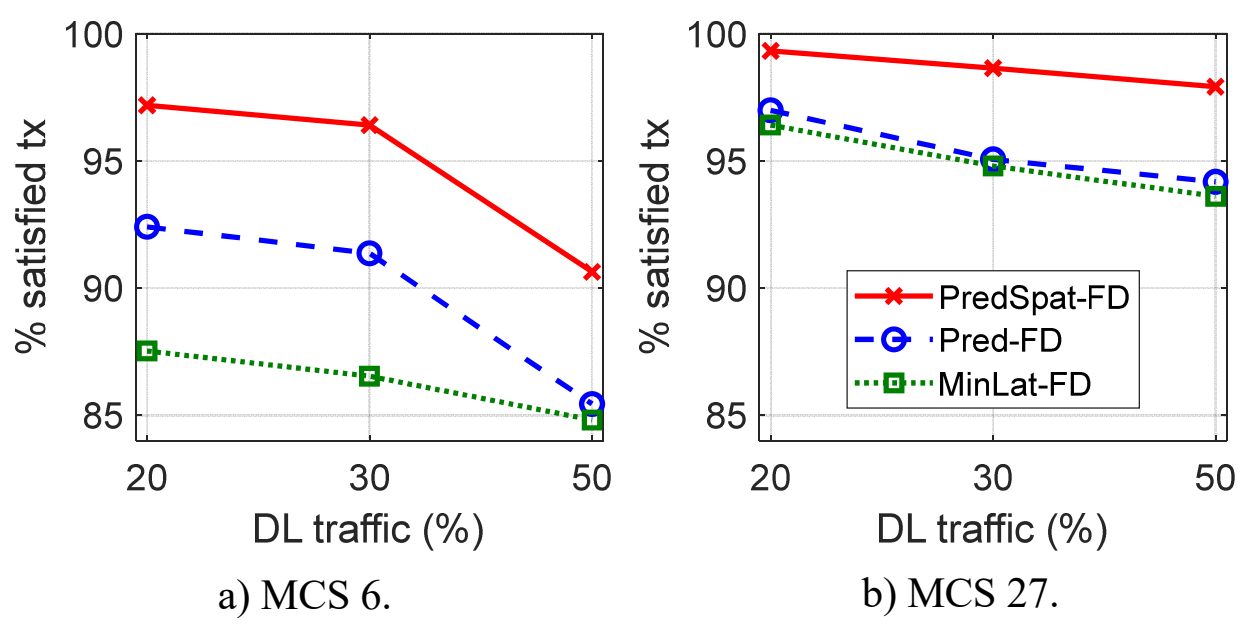

a) MCS 6. b) MCS 27.

Fig. 8. Percentage of satisfied transmissions with varying DL traffic. (Traffic configuration C2).

conflicts compared to Pred-FD by selecting predicted regions associated with spatially distant nodes.

Despite Pred-FD and PredSpat-FD experience similar UL/DL conflicts (Fig. 5), especially under C1, Fig. 4 shows higher percentage of satisfied closed-loop sessions achieved by PredSpat-FD. This is because of the spatial diversity differentiation criterion between them. When UL/DL conflict-free transmissions cannot be scheduled, PredSpat-FD seeks UL/DL conflicts between SNEs that are highly separated (i.e., experience low CLI). For example, with PredSpat-FD, DL transmissions in conflict with UL transmissions from another closed-loop session experience on average +5.6 dB SINR increase compared to Pred-FD. This result, shown in Fig. 6 for MCS index 11, is consistent across both C1 and C2 and for other evaluated scenarios. This SINR increase leads to a higher ratio of conflicted DL transmissions that can be correctly received by the proposed PredSpat-FD scheme (Fig. 7). As shown in Fig. 7, the two reference schemes are challenged to satisfy their conflicted DL transmissions at higher MCS indices due to their lower tolerance to interference (i.e., they require higher SINR levels). In contrast, the spatial diversity policy followed by the proposed scheme when simultaneously scheduling UL and DL transmissions mitigates CLI, resulting in higher SINR levels (Fig. 6), and consequently, greater support for correctly receiving conflicted UL/DL transmissions even at higher MCS levels. Finally, Fig. 8 evaluates the scalability of the scheduling schemes under different MCS and DL traffic configurations. The results show that PredSpat-FD achieves the highest percentage of satisfied closed-loop sessions across all evaluated scenarios. Unlike MinLat-FD, whose performance degrades as DL traffic increases, and Pred-FD, which provides limited gains due to the absence of spatial diversity, PredSpat-FD sustains high performance even at higher loads. The fallback mechanism is rarely triggered ($\lesssim$1.25% for MCS 6 under high DL size) and negligible at higher MCS values, indicating minimal impact on overall performance. These results demonstrate that the joint use of traffic prediction and spatial diversity within FD enables the support of time-critical closed-loop applications, even under high network loads.

## V. Conclusions

We presented in this paper a novel scheduling scheme that combines traffic prediction and spatial diversity to address the inherent CLI of FD. The scheme is evaluated in the context of a realistic 6G in-vehicle subnetwork supporting a mix of time-sensitive closed-loop traffic with varying latency requirements. Our evaluations show that the proposed PredSpat-FD scheme reduces UL/DL conflicts that cause CLI and increases SINR when conflicts are unavoidable. As a result, it achieves a higher rate of closed-loop sessions completed within their deadlines compared to reference schemes that do not incorporate spatial diversity and predictive scheduling.